\documentclass[twocolumn,showpacs]{revtex4}
\usepackage{graphicx}

\begin{document}

\title {Effect of dielectric coating on the positron work function of a metal}

\author {V. V. Pogosov,  A. V. Babich}

\address {Department of Micro- and Nanoelectronics,
Zaporozhye National Technical University, Zhukovsky Str. 64,
Zaporozhye 69063, Ukraine}

\date {\today}

\begin {abstract}
We show that the dielectric coating of the metal surface leads to
the change in the sign of the positron work function.

\end {abstract}


\maketitle

Positron work function is of importance for a proper description of
both bulk and surface states in a metal.

Electron work function $W_{\rm e^{-}}$ for metals is always positive
(metal can be considered as a potential box for ${\rm e^{-}}$). For
some metals, positron work function $W_{\rm e^{+}}$ has a negative
sign \cite{121,122} (these metals are equivalent to the potential
barrier for emitted ${\rm e^{+}}$). Among the metals with a negative
positron work function are cooper, aluminium, iron, molybdenum,
nickel, chromium, and titanium. For these metals, the value of
$W_{\rm e^{+}}<0$ was obtained from reemitted positron energy
distribution relatively to the vacuum level (see Refs.
\cite{121,122,1291} and references therein).

Electron and positron work functions are controlled by their bulk
contributions as well as by the surface \emph{electrostatic} dipole
barrier $D$, which enables one to distinguish between different
crystal faces:
\begin{equation}
W_{\rm e^{-}}=W_{\rm e^{-}}^{\rm bulk}+eD, \quad W_{\rm
e^{+}}=W_{\rm e^{+}}^{\rm bulk}-eD,\label{WE+WP}
\end{equation}
where $e$ is the unit positive charge and the term $W_{\rm
e^{+}}^{\rm bulk}$ includes the positron band shift energy and
positron-electron correlation contribution. In the present paper,
for Al, Cu, and Zn, we use the following values of $W_{\rm
e^{+}}^{\rm bulk}: 3.97,\, 2.82$ and 3.80 eV, respectively, these
values being extracted from the literature (see Ref. \cite{122} and
references therein).

By assuming that the value of the electrostatic potential far beyond
the metal is equal to zero it is possible to find the dipole barrier
$D=-\bar{\phi}$, where  $\bar{\phi}<0$ is the bulk value of the
electrostatic potential \cite{Bab-2008}. Owing to inequalities
$W_{\rm e^{+}}^{\rm bulk},\,D>0$, the strong competition between the
terms in Eq. (\ref{WE+WP}) leads to the negative sign of $W_{\rm
e^{+}}$.

In Ref. \cite{Bab-2008}, we reported on the method to calculate
energy characteristics of metallic surface (including $D$) covered
by an insulator coating as functions of the dielectric constant
$\varepsilon$. The calculations are performed by using the Kohn-Sham
method and stabilized jellium model. Strong dependence of $D$ on
$\varepsilon$ allows one to establish that the sign of $W_{\rm
e^{+}}$ can be changed  by changing $\varepsilon$ of coating, while
keeping the same metal (or the crystal face).

\begin{figure}
\centering
\smallskip\smallskip
\includegraphics [width=.4\textwidth] {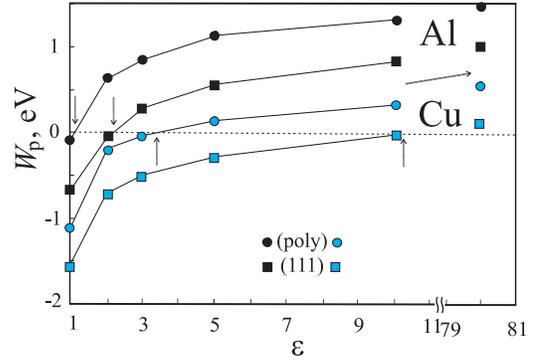}
\caption{The calculated dependence $W_{\rm
e^{+}}(\varepsilon)$.\smallskip} \label{W(eps)}
\end{figure}

The effect of sign for Al è Cu is illustrated in Fig. \ref{W(eps)}.
For polycrystalline  Zn / Zn (0001), the effect of $\varepsilon$
does not result in any dramatic changes: $W_{\rm p}=+0.61\,/-2.59$
and $+2.01\,/-1.23$ eV for $\varepsilon=1$ and 80, respectively.
Similar conclusions can be drawn for Al (110) / (100) [$W_{\rm
p}=-4.86\,/-2.23;\,-3.39\,/-0.55$ eV] and Cu (110) / (100) [$W_{\rm
p}=-4.52\,/-2.64;\,-3.02\,/-0.95$ eV ] at $\varepsilon=1;\,80$,
respectively.

Note that the positronium work function can be calculated from the
Born cycle,
$$
W_{\rm Ps}=W_{\rm e^{-}}+W_{\rm e^{+}}-{\rm Ry/2}.
$$
$W_{\rm Ps}$ does not depend on $D(\varepsilon)$ but it is
determined entirely by the bulk properties.

To conclude, we have demonstrated that the sign of the positron work
function is sensitive to the dielectric coating of the metal
surface. The reported results on the work function sign change can
be of importance for applications in positron diagnostics of
adsorbates or oxide at surface  and possibly in nanotechnologies.

\end{document}